\documentclass{iau} 
\usepackage{graphicx}
\title[Populations of accreting WDs] 
{Populations of accreting white dwarfs}

\author[Chen et al.]   
{Hai-Liang Chen,$^{1,2,3}$
 Tyrone E. Woods,$^4$
 Lev Yungelson,$^5$
 Marat Gilfanov,$^{6,7}$
 \and Zhanwen Han$^{1,2,3}$}

\affiliation{
$^{1}$Yunnan Observatories, CAS, Kunming, 650011,China\\ email:{\tt chenhl@ynao.ac.cn}\\
$^{2}$Key Laboratory for the Structure and Evolution of Celestial Objects, CAS, Kunming 650011, China\\
$^{3}$Center for Astronomical Mega-Science, Chinese Academy of Science, Beijing 100012, China\\
$^{4}$Institute of Gravitational Wave Astronomy and School of Physics and Astronomy, University of Birmingham, Birmingham B15 2TT, UK\\
$^{5}$Institute of Astronomy, RAS, 48 Pyatnitskaya Str., 119017 Moscow, Russia\\
$^{6}$Max Planck Institute for Astrophysics, Karl-Schwarzschild-Str. 1, Garching b. M{\" u}nchen 85741, Germany\\
$^{7}$Space Research Institute of Russian Academy of Sciences, Profsoyuznaya 84/32, 117997 Moscow, Russia\\
}    
    
\pubyear{2018}
\volume{xxx}  
\jname{Title of your IAU Symposium}
\editors{A.C. Editor, B.D. Editor \& C.E. Editor, eds.}
\begin{document}

\maketitle

\begin{abstract}
 Using a hybrid binary population synthesis approach, we modelled the formation and evolution of populations of
accreting WDs for differing star formation histories. We found that the delay time distribution of SNe Ia 
in the single degenerate scenario is inconsistent with observations. Additionally, we found that our predicted X-ray and UV 
emission of populations of accreting WDs are consistent with the X-ray luminosities of early-type galaxies observed 
by Chandra and the HeII 4686$\AA$/H$\beta$ line ratio measured in stacked SDSS spectra of passively evolving galaxies. 
Moreover, we found that the majority of current novae in elliptical-like galaxies                     
have low-mass WDs, long decay times, long recurrence periods and are relatively faint. 
In contrast, the majority of current novae in spiral-like galaxies have massive WDs, 
short decay times, short recurrence periods and are relatively bright. Our predicted distribution of mass-loss timescales in an M31-like galaxy is consistent with observations for Andromeda.

\end{abstract}

\firstsection 
\section{Introduction}
It is now understood (e.g., \cite{Nomoto07}) that for a narrow range of accretion rates, hydrogen will undergo steady nuclear-burning on the surface of an accreting white dwarf.
 If the accretion rate is below this range, hydrogen burns unstably, observed as novae. In this regime, accreting WDs produce little X-ray emission.
In the stable burning regime,  accreting WDs have a typical effective
temperature of $10^{5}-10^{6}$K, and radiate predominantly in the soft X-ray and EUV bands.
For accretion rates above the stable burning regime, the evolution of accreting WDs is still unclear. In one scenario, suggested by
\cite[Hachisu et al. (1996)]{hkn96} an optically thick wind is launched. In this case, the typical effective temperature 
of accreting WDs in the wind regime is $10^{4}-10^{5}$K and WDs emit prominently in the EUV. 
In this work, we modelled the formation and evolution of accreting WDs 
in different types of galaxies using a hybrid binary population synthesis approach. We 
study their X-ray and UV emission and properties of the nova population. 

\section{Hybrid Binary Population Synthesis Approach}

Our calculations consist of two steps. First, we use the \textsc{bse}  code (\cite{htp02}) to obtain the population of WD
binaries with non-degenerate companions at the onset of mass transfer. Then we use the \textsc{mesa} (\cite{pbdh+11}) code
to compute a grid of models describing the evolution of  WD binaries with varying initial parameters. 
With the binary parameters at the onset of mass transfer from our \textsc{bse} calculations, we can 
select the closest track in the grid of \textsc{mesa} calculations and follow the evolution of any WD binary.
For more details, we refer to \cite{cwyg+14}.

\section{Results}

In \cite{cwyg+14}, we have computed the SN Ia rate in the single degenerate scenario in different types of galaxies, 
and compared these results with observational constraints.
We found that the delay time distribution of SNe Ia in our calculation is inconsistent with observations (see Fig.~8 in \cite{cwyg+14}), being more than 10 times smaller than the observationally inferred value.

In \cite{cwyg+15}, we computed the X-ray luminosity in the soft X-ray (0.3-0.7 keV) band for elliptical-like galaxies as a function of stellar age and 
found that it is comparable to the Chandra observational data of nearby elliptical galaxies (see Fig.~9 in \cite{cwyg+15}). In addition, 
we computed the time evolution of the He II $\lambda 4686$/H$\beta$ line ratio in passively evolving galaxies, and found that it is in good agreement with the line ratio 
measured in stacked SDSS spectra of retired galaxies (Johansson et al. 2014, 2016;  Fig.~11 in \cite{cwyg+15}).  

In \cite{cwyg+16}, we have modelled the evolution of the nova population in different types of galaxies. We found that the current nova rate per unit mass in 
elliptical-like galaxies is 10-20 times smaller than that in spiral-like galaxies. Moreover, we found that the current nova population 
in elliptical-like galaxies has lower-mass WDs, longer decay times, relatively fainter absolute magnitudes and longer recurrence periods.
The current nova population in spiral-like galaxies have massive WDs, short decay times, are relatively bright and have short 
recurrence periods. In addition, the predicted distribution of mass-loss timescale in a M31-like galaxy is in good agreement with observed statistics (see Fig.~ 3 in \cite{cwyg+16}).

\section{Acknowledgements}
This work is partially supported by the National Natural Science Foundation
of China (Grant no. 11703081,11521303,11733008), Yunnan Province (No. 2017HC018) and the CAS light of West China Program.


\end{document}